
\input harvmac

\def \ra {\rightarrow}
\def \na {\nabla }
\def \D {\Delta}
\def \tD {{\tilde{\Delta}}}
\def \bZ {{\bar{ Z}}}
\def \tQ {{\tilde{ Q}}}
\def \ty { {\tilde y}}
\def \by { {\bar  y}}
\def \tG { {\tilde G}}
\def \tx { {\tilde x}}

\def \hi {\chi}

\def \Tr {{\rm Tr \ }}
\def \log {{\rm log \ }}

\def \det {{\rm det \ }}

\def \l {\lambda}
\def \4p {{1\over 4 \pi }}
\def \8p {{1\over 8 \pi }}
\def \P^* { P^{\dag } }
\def \p {\phi}
\def \M {{\cal M}}
\def \m {\mu }
\def \n {\nu}
\def \ep {\epsilon}

\def \r {\rho}
\def \k {\xi}
\def \d {\delta}
\def \o {\omega}
\def \s {\sigma}

\def \fourth {{1\over 4}}
\def \third {{1\over 3}}
\def \six {{1\over 6}}
\def \e#1 {{\rm e}^{#1}}
\def \const {{\rm const }}

\def \eq#1 {\eqno {(#1)}}
\def\np {  Nucl. Phys. }
\def \pl { Phys. Lett. }
\def \mpl { Mod. Phys. Lett. }

\def \ap  { Ann. Phys. }
\def \cmp { Commun. Math. Phys. }

\Title{Imperial/TP/92-93/01
\ \ \ hep-th/9210015}
{\vbox{\centerline{ Dilaton shift under duality}
\vskip2pt\centerline{ and torsion of  elliptic complex}}}
\centerline {A.S. Schwarz
\footnote {$^\dagger$}
{e-mail: asschwarz@ucdavis.edu}}
\centerline {\it Department of Mathematics, University of California }
\centerline {\it Davis, CA 95616, USA}

\bigskip
\centerline{ A.A. Tseytlin
\footnote{$^*$}
{e-mail: aat11@amtp.cam.ac.uk
}
\footnote{$^\star$} {On leave of absence from the Department of
Theoretical Physics, P. N. Lebedev Physics Institute, Moscow 117924, Russia.}}
\centerline{\it  Theoretical Physics Group }
\centerline {\it  Blackett Laboratory}
\centerline{\it Imperial College}
\centerline{\it  London SW7 2BZ, United Kingdom }

\baselineskip=20pt plus 2pt minus 2pt
\vskip .3in

We observe that the ratio of  determinants of $2d$ Laplacians
which appear  in the duality transformation relating two sigma models  with
abelian isometries can be represented as a torsion of an elliptic
(DeRham) complex. As a result, this ratio  can be computed exactly and is
given by the exponential of  a local functional of $2d$ metric and target
space metric. In this way the well known dilaton shift under duality is
reproduced. We also present the exact computation of  the determinant which
appears  in  the  duality transformation   in  the   path  integral.

\Date{9/92} 


\newsec{Introduction}
Duality transformations in two dimensional sigma models with abelian
isometries have recently attracted much attention in connection with  string
theory (see e.g. [1-4] and refs. there).\foot{For duality transformations in
the
non-abelian case see e.g.[5].} As is well known [1,6,3], the duality
transformation which inverts  the metric in the direction of an  isometry
should also include  a shift of the dilaton field (coupled to the
curvature of the $2d$ metric).  Let us consider a sigma model
$$I= {1\over { 4 \pi  }} \int d^2 z \sqrt {g} \ [ (G_{\m \n} + B_{\m \n}) (
g^{ab} + i \ep^{ab}) \del_a x^\m \del_b x^\n \  +  \  R  \p \ ] \  \ ,
\eq{1} $$
which is invariant under a global isometry $\d x^\m = \ep K^\m $. Then $K^\m$
must be  a Killing vector of $G_{\m \n }$ and the Lie derivatives of $B_{\m \n
}
$ (and $\p$ ) must vanish up to a gauge transformation term. In general, one
can
choose local coordinates  $ \{ x^\m \} = \{ x^1 \equiv y \ , \  x^i \} $ such
that $G_{\m \n}$, $B_{\m \n}$  and $ \p $ are independent of  the coordinate
$y$
which is shifted by the isometry. Then the action (1)  is at most
quadratic in $y(z)$.  To obtain the dual model one  replaces $\del_a y$
by a ``momentum" field $p_a$ adding the constraint term $ \sim \ty \ep^{ab}
(\del_a p_b - \del_a p_b ) ,$ where $\ty$ is  a Lagrange multiplier.
 By formally doing  the gaussian  integral over $p_a$ one finds  an  action
(dual action $\tilde {I}$)
for  $ \{ \tx^\m \} = \{ \tx^1 \equiv \ty \ , \  x^i \} $
 which has the form (1)  with
$$ {\tilde G}_{11} = M^{-1} \ \ , \ \ {\tilde G}_{1i} =  M^{-1} B_{1i} \ \
, \ \ {\tilde G}_{ij} =  G_{ij} - M^{-1} ( G_{1i} G_{1j}- B_{1i} B_{1j})\ \ ,
$$ $$
\ \ {\tilde B}_{1i} =  M^{-1} G_{1i} \ \
, \ \ {\tilde B}_{ij} =  B_{ij} - M^{-1} ( B_{1i} G_{1j}- G_{1i} B_{1j})\ \ ,
\ \ \ M \equiv G_{11} \ \ . \ \eq{2} $$
The integral over $p_a$ is, however, ill-defined  given that   the
coefficient $M(x)$ of the $p^2$ term is  just a local function and not an
elliptic differential operator. Therefore one cannot assert that the
 partition   functions corresponding to action functionals $I$ and $\tilde {I}$
are equal. However one may expect that the ratio of  these partition
functions should be given by the exponential  of a local functional $$  \ \
\Delta I =  {1\over { 4 \pi  }} \int d^2 z \sqrt {g} \ [
 c_0 \e{2\l} +  c_1
\del_a \l \del^a \l  \  +  \  c_2  R \l \ ] \ \ , \ \   M \equiv \e{2\l}
\ \ . \eq{3} $$
The coefficients $c_n$ should depend on a particular way of defining the
integral
over  $p_a$.  The definition  suggested  in [1]   was based on
changing the  variables
 $$ p_a = \del_a \k^1 + \ep_{a}^{ \ b } \del_b \k^2  \ \ , \eq{4} $$
and  computing  the  determinant of the  corresponding second order
differential operator $Q$  defined on  the scalar fields $ \k^n$.  The
part of this determinant which depends on the conformal factor of the $2d$
metric $g_{ab}$  is determined by the Weyl anomaly  and can be easily extracted
[1,3]. As a result, one finds  that $c_2$ in (3) is equal to --1,  implying
that the dilaton  should be  shifted under the duality by $-\ha \log M $.
There
remains, however, the question about other possible   terms in the determinant
of $Q$ which are Weyl invariant but $M$ - dependent. It is not  clear  why
such terms  in $\det Q $ should be local.
 This question  was not resolved in [1,3].\foot  {It was noted in [3] that
$\det
Q$ may be exactly computable using the standard  variational argument
but a consistent derivation was not carried out.} We shall present the
exact  computation of $\log \det Q$ in Appendix and show that it is given by a
local expression  (3) (with non-vanishing $c_1 $).  The $\del_a \l \del^a
\l$ term  coming from $\log \det Q $ is cancelled by the contribution of
another
 determinant (ignored in [1]) which appears in the process of duality
transformation in the path integral.

 In general, one would like to fix the structure of finite local counterterms
(3) in such a way that the original and dual sigma  models  remain equivalent
at
the quantum level. The necessary condition  for the equivalence  is  that (3)
should compensate for the non-trivial  ratio of the determinants which appear
as a result of  integrating  over $y$ and $\ty$ in the original and dual
theories, $$ Z = { { \int [dy] \exp [-\int d^2z \sqrt g M(x) \del_a y \del^a y
]
} \over { \int [d\ty] \exp [-\int d^2z \sqrt g M^{-1}(x) \del_a \ty \del^a \ty
]
} }  \ \ ,\eq{5} $$  Ignoring the zero mode factors $$
Z =[\det \D_0 ]^{-1/ 2} [\det \tD_0 ]^{1/2}  \ \ .  \ \
\eq{6} $$
 However, since each of
the determinants in (6)  is a complicated non-local functional  of the function
$M$ it is not {\it a priori } clear why the logarithm of their ratio (6) should
have a local form.  In Sec.3  below we shall   prove that (6) is, in fact,
given by a local expression   which structure  in general depends on choices of
the  scalar products in the corresponding spaces of $y$ and $\ty$.
For the natural choice  of
scalar products we will find that
$$ Z = \exp ({1\over { 8 \pi  }} \int d^2 z \sqrt {g} \
 R\  \log M ) \ \  \ \ , \eq{7} $$
implying that the dilaton is shifted under the duality transformation [1]
$$ {\tilde \phi } = \phi - \ha \ \log M  \ \ . \eq{2'} $$
The proof will  be based  on  the interpretation of  the partition function
(5),(6) as the torsion of the elliptic complex of  exterior differentials
acting on forms (the DeRham complex) in $d=2$ in the case of non-trivial
 scalar products  in the corresponding vector spaces. The result
may be considered as a generalisation of the standard expression
which  gives  the index of the DeRham complex in terms of the Euler number (see
e.g. [7]).

\newsec{ Basic definitions and relations}
We shall start with summarising some relevant information (following  ref.[8])
about  a special combination of determinants
of elliptic operators  which   is  known as   the  torsion of an elliptic
complex.  Let us consider a complex $(E_i,T_i) \ ( i=0, 1,...,N \ , E_{-1} =
E_{N+1} =0$), i.e. a sequence of vector spaces $E_i$ and linear operators $T_i$
acting from the space $E_i$ to the space $E_{i+1}$  and satisfying $ T_{i+1}
T_i
=0$. Let $<\ ,\ >_i $ denote a scalar product in $E_i$ and
define the adjoint operators $T^*_i: E_{i+1} \ra E_{i} $ by $ <a,T_ib>_{i+1} =
<
T^*_ia,b>_{i}$. If the self-adjoint operators $\D_i: E_{i} \ra E_{i} $
 $$ \D_i
= T^*_i T_i +  T_{i-1}T^*_{i-1} \ \ \eq{8} $$ are elliptic differential
operators  the complex $(E_i,T_i)$ is called  an elliptic  complex. The
torsion
$Z(E_i,T_i)$  of the elliptic complex $(E_i,T_i)$ is  given  by the
formula\foot {
Note that  we have changed the notation as compared to ref.[8]:
our $T_i$ is $T_{N-i}$ of [8].}
$$  \log Z(E_i,T_i) =  \ha \sum_{i=0}^{N} (-1)^{i}  (i+1)\ \log \det \D_i \ \
, \  \eq{9} $$
where the regularised determinants of operators $\D_i$  are defined  by means
of $\zeta$-function. $Z$ can be identified with the partition function of a
 degenerate functional (associated with  $(E_i, \ T_i)$) with zero action [8].

Using that $$ T_{i+1}
T_i =0  \ \ , \ \ \ \det \D_i = \det (T^*_i T_i ) \ \det (T_{i-1} T^*_{i-1} ) \
\
, \ \ \  \det (T_i T^*_i )=  \det (T^*_{i} T_{i} ) \ \ , \eq{10} $$
one can represent $Z$ in the form
$$  Z= \prod_{i=0}^{N}  (\det T^*_i T_i)^{\ha (-1)^{i+1} }
 \ \ . \eq{11} $$
Since $T^*_i$ depend on a definition of the scalar products in $E_i\ ,$
$Z$ changes under a  variation of the scalar products. If one redefines the
scalar products by inserting the operators $M_i: E_i \ra E_i \ , $
$$ < \ , \  >_i'= < M_i \ ,   \ >_i \ \ , \ \
  \ \ \ {T^*_i}{}'= M_{i}^{-1} T^*_i M_{i+1} \ \  \ . \eq{12}  $$
As was shown in [8], the variation of the partition function (9),(11) under the
variation  of the scalar products in $E_i$
can be expressed in terms of the  Seeley coefficients (i.e. local functionals
appearing in  the  $t\ra 0$ asymptotics of $\exp (-t\D_i) $)  and the zero
modes of $\D_i$.
If $$ {\d} \l_i = \ha M^{-1}_i \d M_i$$ are the operators  describing
an infinitesimal change of the scalar products,\foot {The proof of this
formula is based on representing the determinants in terms of proper-time
integrals of the heat kernels and relating the variations of heat kernels
corresponding to different  operators.}
 $$ \d \ \log Z =  \sum_{i=0}^{N} (-1)^i [ \Psi_0 ({\d} \l_i\vert
\D_i) -
 P ({\d} \l_i\vert \D_i)]  \  \ .  \eq{13} $$
Here $\Psi_0$ is the coefficient of $t^0$ in the asymptotic expansion
$$ {\rm Tr }({\d } \l_i { \e{-t\D } }) \sim \sum_k \Psi_k({\d}
 \l_i \vert \D_i)\  t^k \ \ , \eq{14} $$
and
$$ P({\d} \l_i\vert \D_i) = \sum_n <{\d} \l_i f^{(n)}_i,   f^{(n)}_i >_i \ \ ,
\eq{15} $$
where $ f^{(n)}_i$  is a basis in the kernel of $\D_i$.

Let us now specialise to the case when $(E_i, T_i)$ is the DeRham  complex,
i.e.  when $E_i$ are the spaces of $i$-forms on a compact
 riemannian space $\M^{N}$ with the metric $g_{ab}$ and $T_i=d_i$ are
exterior differentials. The torsion (9),(11)
can be considered as the  partition function  of the
quantum theory  of the antisymmetric tensor of rank $N$  which has zero action
[8--11]. We are interested in the generalisation of the standard discussion to
the case when the scalar products in $E_i$ are non-trivial, namely  contain
additional  scalar functions $M_i (z)$ $$ < \o, \s >_i = \int d^{N}z { \sqrt g
}
\ M_i g^{a_1b_1}... g^{a_ib_i}\ \o_{a_1 ... a_i}  \s_{b_1 ... b_i}\ \ . \eq{16}
$$ Then according to (12), (8)
$$ T^*_i= M_{i}^{-1} d^*_i M_{i+1} \ \  \ , \eq{17}  $$
$$ \D_i =  M_{i}^{-1} d^*_i M_{i+1} d_i +  d_{i-1} M_{i-1}^{-1}
d^*_{i-1}M_{i}\ \ . \eq{18} $$
When $M_i=1$  the partition function $Z$ (11)  is known as the Ray-Singer
torsion $tor \M $ of the manifold $\M^{N}$ [9,11]  the variation of which under
a  change  of  the metric $g_{ab}$  can be expressed in terms of  the
 ``anomalies" and zero modes of $\D_i$ ($tor \M $ is equal to 1 if $N$ is
even).  Note that $ \det \D_i \not= \det \D_{N-i} $ when $M_i\not=1$.

 The variation of $Z$ (13) under a variation of $M_i$ takes the form
 $$ \d \ \log Z =  \sum_{i=0}^{N} (-1)^i \ [\ \int d^{N} z  \ \d \l_i \
b_{N} (z\vert \D_i) -  \sum_n <{\d} \l_i f^{(n)}_i , f^{(n)}_i>_i \ ]
  \  \ ,   \eq{19} $$ $$ \ \ \ M_i \equiv \e{2 \l_i} \ \ . $$
Here $b_{N}(z\vert\D_i)$ is the local Seeley  coefficient (integrand of
$\Psi_0$) of  the Laplacian
 $\D_i$  and  $f^{(n)}_i$ are the zero modes of $\D_i$ (for a
discussion of the zero mode contribution see [11]).  Note that for $M_i=1$
$$ \sum_{i=0}^{N} (-1)^i \int d^{N} z \  b_{N}
(z \vert \D_i) $$
is the index of the DeRham complex  which by index theorem is equal to the
Euler
number $\hi$ of  $\M^{N}$.
Eq.(19) then implies that (up to the contribution of the zero modes) the  power
of the common  constant scale of $M_i$ in $Z$ is equal to  $  \ha \hi  $. If
$b_{N}(z\vert\D_i)$  and $f^{(n)}_i$  are  known explicitly, eq.(19) gives a
system of functional  equations $$ {{\d \log Z }\over \d \l_i} =
F(\l_1,...,\l_{N}; g_{ab}) $$   which, in principle, can be integrated to find
the dependence of $Z$ on $M_i$.

\newsec{ Torsion of generalised DeRham complex in two dimensions}
The case of our interest is when  the complex is defined on a
two-dimensional Riemann space, i.e.  the simplest non-trivial case of the
above construction ($N=2$),
$$  0 \ra E_0 \ra E_1 \ra E_2 \ra 0 \ \ . $$
$d_0$ acts from the space of scalars $E_0$ to the space of vectors (1-forms)
$E_1$  and $d_1$ acts from $E_1$ to the space of 2-forms $E_2$. In two
dimensions  2-forms can be identified with (``dual") scalars ($\o_{ab}=
\ep_{ab}
\o$). The  corresponding scalar products and Laplacians  are given by (16),(18)
 $$ < \o , \s >_0 = \int d^{2}z { \sqrt g } \ M_0 \ \o \s \ \ , \ \
< \o , \s >_1 = \int d^{2}z { \sqrt g } \ M_1 \ g^{ab} \o_a \s_b \ \ , \ \
$$ $$ < \o , \s >_2 = \int d^{2}z { \sqrt g } \ M_2\ g^{ab}g^{cd} \ \o_{ac}
\s_{bd} \ \ , \ \  , \eq{20} $$
$$ \D_0 = M_{0}^{-1} d^*_0 M_{1} d_0  \ \ , \ \ \
\ \D_1 =  M_{1}^{-1} d^*_1 M_{2} d_1 +  d_{0} M_{0}^{-1}
d^*_{0}M_{1}\ \ , \ \ \
\D_2 = d_1 M_{1}^{-1} d^*_1M_{2}  \ \ , $$
$$\D_0= -  M_{0}^{-1} \na^a (M_{1} \na_a ) \ \ ,\ \ \ \  \D_2 = -  \na^a
(M_{1}^{-1} \na_a M_{2})  \ \
$$ $$
 \ \ \ \D_{1ab}= -  M_{1}^{-1} \na^c (M_{2} ( g_{ab}\na_c
-g_{cb}\na_a ))  - \na_a (M_{0}^{-1} \na_b M_{1})  \ \
 . \ \ \eq{21} $$
 Since the determinants  are invariant under $\D \ra S^{-1} \D S $ where
$S$ is an operator of multiplication by a function $\D_2$ can be represented
also
in the equivalent form
 $$\D_2 = -  M_{2} \na^a (M_{1}^{-1} \na_a )  \ \ . \eq{21'}  $$
The partition function $Z$ in (9),(11)  is given by
 $$  Z=
[\det \D_0]^{1/2}  [\det \D_1]^{-1}  [\det \D_2]^{3/2} \ \ \ \eq{22} $$
$$ = [\det \D_2]^{-1/2}  \det \D_1  [\det \D_0]^{-3/2}
$$ or
 $$Z= [\det \D_0]^{-1/2}  [\det \D_2]^{1/2} \ \ . \eq{23} $$
Comparing  (22),(23)  with  (5),(6) we conclude that $Z$ in (23)   gives the
well-defined expression for  the ratio of the partition functions of the two
dual scalar theories  in the general case of   arbitrary functions in the
scalar products ($M=M_1$). Our aim will be to compute $Z$ explicitly as a
functional of $M_i(z)$ and $g_{ab}$ using the variational relation (19). Let us
first ignore the contribution of the zero modes. Then (19)  takes the form
 $$ \d \ \log Z =  \int d^{2} z  \ [\ \d \l_0 b_2 (z\vert \D_0)
-\d \l_1 b_2 (z\vert \D_1) +\d \l_2 b_2 (z\vert \D_2)\ ]
  \  \ .  \eq{24} $$
The variation of $Z$ under the Weyl rescaling of the metric $ \d g_{ab} = 2
\d \r g_{ab}$ can be represented as a particular case of (24) with $ \d \l_0=
\d \r \ , \ \ \d \l_1=0\ , \ \ \d \l_2 = - \d \r $, i.e.
 $$ \d \ \log Z =  \int d^{2} z  \ \d \r \ [ \ b_2 (z\vert \D_0)
- b_2 (z\vert \D_2) \ ]  \  \ .  \eq{25} $$
This expression, being the difference of the conformal anomalies of the scalar
theory and its dual, is  also an obvious  consequence of (23).

 We shall use the  following standard result for the Seeley
 coefficients   of the  Laplace-type operators (21) in two
dimensions [12].  Consider the elliptic  differential operator
$$ \D = -  I g^{ab} \del_a \del_b  - 2 A^a \del_a +  Y \ \ , \ \eq{26} $$
where  $a,b=1,2$, $g^{ab}(z) $ is positive definite,  $I$ is $n\times n$
identity
matrix and $A^a$ and $Y$ are  $n\times n$  matrix valued functions in $R^2$.
Then
$$ b_2 (z\vert \D) = {1\over 4 \pi }\sqrt g  {\rm Tr} ( \six R I -  \na_a A^a
- g_{ab} A^aA^b - Y )  \ \ , \eq{27} $$
where $g_{ab}$ is the inverse of $g^{ab}$, $R$ is the curvature of $g_{ab}$
and $ \na_a A^a = {1 \over {\sqrt g }} \del_a ( {\sqrt g } A^a) $.  The
operator
(26) can be represented also  as
$$ \D=  -  g^{ab} D_a D_b   + X  \ \ , \ \eq{28} $$
where the covariant derivative $D_a$ contains both the Christoffel and
 $A_a$ connection terms.  The equivalent form of (27) is\foot { Note that
$b_2$ is invariant under the similarity transformation $\D \ra S^{-1} \D S ,$
where $S$ is the operator of multiplication by function.}
  $$ b_2 (z\vert \D) =
{1\over 4 \pi }\sqrt g  \ {\rm Tr}\  ( \six I R  -  X)  \ \ . \eq{29} $$
 The operators $\D_0$ and $\D_2$  in (21) can be represented in the form (26)
(with the metric rescaled by  $M_1/ M_0$ and $M_2/M_1$). Applying (27)  and
returning back to the original metric in (21) we find
  $$ b_2 (z\vert \D_0) = {1\over 4 \pi }\sqrt g \  [ \six R   + \third \na^2
(\l_1 -\l_0) - \na^2 \l_1   - \del_a \l_1\del^a \l_1]  \ \ , \ \ \ M_i
\equiv \e{2 \l_i} \ \ , \eq{30} $$
  $$ b_2 (z\vert \D_2) = {1\over 4 \pi }\sqrt g \ [ \six R   + \third \na^2
(\l_2 -\l_1)  +  \na^2 \l_1   - \del_a \l_1\del^a \l_1]  \ \ . \eq{31} $$
Substituting (30),(31) into (25) and integrating  the resulting  equation we
find the dependence of $Z$ on the  metric $g_{ab}$\foot { We consider the case
of the  spherical topology so the dependence on the metric is determined by the
dependence on its conformal factor.}
 $$  \log Z = \4p \int d^2z \sqrt g\  [ R \l_1 + \six R
(\l_0 + \l_2 - 2 \l_1)  ] + O(\l_0, \l_1,\l_2) \ \ . \ \eq{32} $$
For  the case of equal $M_i$ ($\l_i=\l$) this expression was found in [3].
To determine the $\l_i$-dependent terms in (32) we need to integrate eq.(24).
Since for general $M_i$ the vector operator $\D_1$ in (21) is not of the type
(26) we shall do this in two steps. First, we substitute (30) and (31) into
(24) and integrate over $\l_0$ and $\l_2$. Taking into account (32) we get
 $$  \log Z = \4p \int d^2z \sqrt g \ [\ R \l_1 + \six R
(\l_0 + \l_2 - 2 \l_1)     $$ $$ +  \six (\l_0- \l_2) \na^2 (2 \l_1 - \l_0 -
\l_2) - (\l_0- \l_2) \na^2  \l_1  - (\l_0+\l_2) \del_a \l_1\del^a \l_1)] +
O(\l_1)   \ \ . \eq{33} $$
To determine the terms which depend only on $\l_1$ we consider the special case
when all $\l_i$ are equal to $\l$. Then $\D_1$ in (21) takes the form
$$ \ \ \D_{1ab}= -  \e{-2\l} \na^c \e{2 \l} ( g_{ab}\na_c
-g_{cb}\na_a )  - \na_a \e{-2 \l} \na_b \e{2\l}  $$ $$ =  - g_{ab} \na^2
  + R_{ab}   -  2 g_{ab} \del^c \l \na_c
  - 2 \na_a \na_b \l  \ \ . \eq{34}
$$ Now (26)--(29) can be  applied and  one finds
$$ b_2 (z\vert \D_1)\vert_{\l_i=\l}
=  {1\over 4 \pi }\sqrt g \  ( \third R  -  R
- 2\del_a \l\del^a \l
)  \ \ . \eq{35} $$
Substituting (30),(31)
and (35) into (24)  and assuming  $\d \l_i = \d \l $ we get
$$ \log Z \vert_{\l_i=\l} =  \4p \int d^2z \sqrt g \ R \l  \ \ . \eq{36} $$
The origin of the ``anomalous" term (36) can be attributed  to the presence of
the $R_{ab}$ term in the vector operator (34) (which gives the $-R$
contribution to (35)).  Comparing (33) to (36) we
 determine  that the $O(\l_1) $ term in (33) must be $ 2 \l_1 \del_a \l_1\del^a
\l_1$, i.e  the final  expression   for $Z$ (up to zero mode factors) is
 $$  \log Z = \4p \int d^2z \sqrt g \ [\ R \l_1 + \six R
(\l_0 + \l_2 - 2 \l_1)    $$ $$  +  \six (\l_0- \l_2) \na^2 (2 \l_1 - \l_0
-\l_2) - (\l_0- \l_2) \na^2  \l_1  - (\l_0+\l_2 - 2\l_1) \del_a \l_1\del^a
\l_1)]
   \ \ . \eq{37} $$
It is easy to check that the derivative of (37) with respect to $\l_1$ is equal
to (35) for $\l_i =\l$.  Note that  in agreement with the index theorem the
common  constant scale of $M_i$  appears in $\log Z$  with the coefficient
equal  to one half of the Euler number.

Let us now  include the contribution of the zero modes.  We shall assume that
$\D_1$ does not have zero modes. The normalised zero modes of $\D_0$
 and $\D_2$  are $$f_0= ( \int
d^2 z \sqrt g \ M_0 )^{-1/2}  \ \ , \ \ \   f_2=  M^{-1}_2( \int d^2 z \sqrt g\
M^{-1}_2  )^{-1/2}  \  \ . \eq{38} $$
Including the corresponding zero mode terms (see (19)) into (24) and
integrating
over $\l_0$ and $\l_2$  we get the  following additional terms  in $Z$ (23)
 $$  ( \log Z )_{z. m. }=  - \ha \log  ( \int d^2 z \sqrt g\  M_0 )
 + \ha \log  ( \int d^2 z \sqrt g \ M^{-1}_2 ) \ \ . \eq{39} $$
As a result, we can represent (23) in  the form
$$  Z
  =  Z'\ (\int d^2 z \sqrt g \ M_0 )^{-1/2}\
 (\int d^2 z \sqrt g\  M_2^{-1} )^{1/2}
\ \ ,  \eq{40} $$
where $Z'$ is the local expression given by  (37).

The  case of equal functions $M_i$  is the one which is  relevant for  the
duality transformation problem discussed  at the beginning of the paper. In
fact,   given the sigma model (1) it is natural to define the scalar product in
the tangent space at $x^\m$ as
$$ < \d x, \d x'> = \int d^2z \sqrt g G_{\m \n } (x) \d x^\m \d {x'}{}^{\n} =
\int d^2z \sqrt g M(x)   \d y \d y' + ...    \ \ . \eq{41} $$
The scalar product corresponding to the dual sigma model (2) is then
$$ < \d \tx , \d \tx'> = \int d^2z \sqrt g \tG_{\m \n } (x) \d \tx^\m \d
{\tx'}{}^{\n} = \int d^2z \sqrt g M^{-1}(x)   \d \ty \d \ty' + ...    \ \ .
\eq{42} $$
The scalar product on scalars $\d y$ is thus determined by the function
$M_0=M$.
As it is clear from  the structure of the quadratic functionals (actions) in
(5),
to be able to identify $M$ with the scalar product function  in the space of
vectors we need to rescale $\ty$ by a factor of $M$, i.e.  we  should identify
$M^{-1} \ty$ with the 2-forms which appeared in the above discussion (so that
$\ty =\const$ corresponds to the zero mode, etc). Then the scalar product on
the
2-forms is also defined by $M$ ( i.e.  all $M_i$ are equal  to $M$) and  the
operators $\D$ and $\tilde D$  in (6) are equivalent to $\D_0$ and $\D_2$.
For $M_i=M$ eq.(40)  takes the form (see (36))
$$   ( \int d^2 z \sqrt g M )^{1/2}\  [\det \D_0]^{-1/2}
$$ $$   =  \exp ({1\over { 8 \pi  }} \int d^2 z \sqrt {g} \
 R \ \log  M ) \ ( \int d^2 z \sqrt g M^{-1} )^{1/2}\  [\det \D_2]^{-1/2}
\ \ .  \eq{43} $$
Note that if $M=\const$ it drops out from (43) (we have assumed that
the 2-space has the topology of a sphere).
 As a result, we have proved that  up to the ratio of the  zero mode
factors $$ {\int dy ( \int d^2 z \sqrt g M )^{1/2}} \over
{\int d\ty ( \int d^2 z \sqrt g M^{-1} )^{1/2}}  \ \ \ \eq{44} $$
(5)  is  indeed given by the local  expression (7).

\newsec {Concluding remarks}
In conclusion, let us make  several  comments.

Eqs.(5),(7),(43) have a straightforward generalisation to the case when the
sigma
model (1) has a number of commuting isometries, i.e. when $y$ and $\ty$ have an
additional index $s$. Then $G_{st} \equiv  M_{st}$ is
 a matrix depending on the rest of the  fields $x^i(z)$ and $M$ in (7),(43)
should be replaced by $\det M $.

The   relation (43) proved   above  may  be useful in trying  to  clarify the
issue of modification of the leading order duality transformations (2),(2$')$
by higher loop effects [3].

Equations (33) and (43) may be of interest  also from  a mathematical point of
view giving the explicit dependence of the torsion of the $N=2$ DeRham complex
 on the scalar products and providing  the ``$M\not= \const$"
generalisation of the  corresponding index theorem. It may be of interest to
study higher  dimensional analogs of (33),(43)
(in particular, in connection with  possible higher dimensional analogs of
sigma model  duality).

The torsion (22),(23) can be  interpreted  as the   partition function of
the
 theory of  rank 2 antisymmetric tensor in two dimensions [8,11]
(i.e. of a `topological' theory  with zero action). What we have found  is that
if non-constant functions are included in the definition  of the scalar
products
the resulting ``effective action" (36),(37)  is a $local$ action of $2d$
gravity
coupled to scalar(s). Thus the $2d$ $scalar-tensor$  gravity  appears as
an `induced' theory.

The methods of [8] and of the present paper may
find applications in other contexts (e.g. in gauged WZW theory) where a careful
account of the dependence on the definition of the  scalar products is
important. The resulting ambiguity in a choice of local counterterms should be
fixed by additional  conditions depending on  particular theory.

\bigskip
\bigskip
\bigskip
We are grateful to the Aspen Center for Physics for the hospitality  while
 part of this work was  done. A.S. was partially supported by
 NSF grant No. DMS-9201366.
A.T. would like to acknowledge    Trinity College, Cambridge
 and SERC for support.

\vfill \eject

\appendix{} {Integration over ``non-dynamical" vector field   and
determinant of operator $ Q$
} 1.  Let us consider  the following  integral over the $2d$ vector field $p_a$
$$ Z_1= \int [dp_a] \exp {[ - \ha \int d^2z \sqrt g \ M_1(z) g^{ab} p_a p_b  }
\ ]\ \ ,
 \eq{A.1} $$
where $M_1$ is a given  function. We are assuming that the measure (scalar
product) for $p_a$ is trivial (if $M_1$  is present also in the scalar product
it
is natural to set $Z_1=1$). This integral is not well defined. Using different
definitions (regularisations) of (A.1) one will get   expressions which
will  differ by local (finite) counterterms.
 A choice of particular definition is dictated by some additional conditions
which the total theory (where (A.1) appears at some intermediate step) should
satisfy.

 Given that $\D_1 = d^*_1 d_1 +
d_0 d^*_0 = ( - g_{ab}\na^2  + R_{ab} )$ is a   Laplace  operator on
vectors one may define (A.1), for example,  as (cf.(34),(35))
$$ Z_1 = \exp [- \ha \Tr (\log M_1 \  \e{ -\ep \D_1 } ) ] $$ $$ =
\exp [ - \ha \int d^2 z \sqrt g\  \log M_1 ( {2 \over \ep }  + \third R - R )]
\ \ . \eq{A.2} $$
It is more natural, however,  to  use  the
 following general  idea: if $M$ is a ``bad" operator one can define its
determinant by introducing an auxiliary operator $P$ such that $P^*MP$ and
$P^*P$ are ``good" operators and setting
$$ \det M \equiv {\det P^*MP \over \det P^*P  } \ \ . \eq{A.3} $$
The result will, in general, $depend$ on a choice of $P$. However the
dependence of $P$ can be calculated; see [8].
  Changing the variables from $p_a$ to a
pair of scalar fields $\k^n$   $$ p_a = \del_a \k^1 + \ep_{a}^{ \ b } \del_b
\k^2  \equiv P_{an} \k^n \ \ , \eq{A.4} $$
 one  can represent (A.1) as
$$ Z_1 = \ J\ Z_Q  \ \ , \ \ \  Z_Q \equiv
[\det Q]^{-1/2}  \ \ ,\ \ \ J= [Z_Q(M_1=1)]^{-1} \ \  ,  \eq{A.5} $$
 where $ J $ is the Jacobian
 of the transformation and  $Q$ is   the following  Laplacian  defined on
a  pair of scalars
 $$  Q\equiv P^{\dag} P \ \ , \ \ \ \ P^{\dag } \equiv M_0^{-1} P^* M_1 \ \ ,
\eq{A.6} $$
$$
Q_{nm}= - M_0^{-1} \d_{nm} \na^a(M_1\del_a) - \ep_{nm} \ep^{ab} M_0^{-1}
\del_a M_1 \del_b  $$
$$ =   \e{ 2(\l_1 -\l_0) } [\ - \d_{nm}\na^2
  -  2( \d_{nm}g^{ab}  + \ep_{nm} \ep^{ab}) \del_a \l_1 \del_b\ ]
 \ \  ,
 \ \  \eq{A.7} $$
$$    \ \  \ M_i \equiv  \e{2 \l_i }  \ \ \ . $$
Here
$$ P^{*na} p_a = ( - \na^a p_a\ , \ \ - \ep^{ab} \na_a p_b ) \ \ , \eq{A.8} $$
 and we  have assumed that  the scalar
product in the space of $\k^n$ is defined  with an extra function $M_0(z)$
(cf.(20)).

Since $Q$ is a $2d$ scalar operator  the dependence of $Z_Q$ on the conformal
factor $\r$ of the 2-metric is  determined by the (Weyl) anomaly.  In view
of the  ``product"  structure of $Q$ the same is  effectively true also for the
dependence on  $M_0$ and $M_1$. Computing the variations of $\det Q$ with
respect
to $M_i$ and $\r$ we get (cf.(24),(25))
 $$ \d \ \log Z_Q =  \int d^{2} z  \ [ \ (\d \r + \d \l_0 )
 b_2 (z\vert Q)  -  \d \l_1  b_2 (z\vert \tQ) \ ]
 \  \ \  . \eq{A.9} $$
Here the operator  $\tQ = PP^{\dag}$ or, equivalently,
 $$ \tQ =
M_1 P M_0^{-1} P^*  \ \ , \eq{A.10} $$ is a Laplacian acting on vectors,
$$ \tQ_{ab}  =
- M_1 \na^c  M_0^{-1} ( g_{ab}\na_c
-g_{cb}\na_a )   - M_1 \na_a  M_0^{-1} \na_b   $$
 $$   =  \e{ 2(\l_1 -\l_0) } [\ - g_{ab}\na^2 + R_{ab}
  +    2( g_{ab}g^{cd}  + \ep_{ab} \ep^{cd}) \del_c \l_0 \na_d \ ]
 \ \  . \eq{A.11} $$
 Note that $ \det {\tilde Q} = \det Q$ and that  the structure of  the
operator $\tQ $ is different from the structure of $\D_1$ in (21).  The Seeley
coefficients   $ b_2 (z\vert Q)$ and  $ b_2 (z\vert \tQ)$   are found using
(26),(27) (cf.(30),(31),(34))
$$  b_2 (z\vert Q)=
{1\over 4 \pi }\sqrt g \  [ \third R  +  {2\over 3} \na^2 (\l_1 -\l_0)
-2\na^2
\l_1 \ ]  \ \ , \eq{A.12} $$
$$  b_2 (z\vert \tQ)=
{1\over 4 \pi }\sqrt g \  [ \third R  - R  +  {2\over 3} \na^2 (\l_1 -\l_0)
+ 2\na^2 \l_0 \ ]  \ \ . \eq{A.13} $$
Integrating (A.9)  we get  (cf.(37))
$$ \log Z_Q (\l_0, \l_1 ) = \4p \int d^2z \sqrt g \ [\ - {1 \over 12 }  R
\na^{-2} R +  R \l_1 $$ $$  -
 \third  R (\l_1 -\l_0)  - \third  (\l_0 -\l_1) \na^2 (\l_0 - \l_1)
  -  2 \l_0 \na^2 \l_1   \ ]
  \ \ , \eq{A.14} $$
$$  \log Z_1 = \log Z_Q (\l_0, \l_1 ) - \log Z_Q (\l_0,  0 ) $$ $$  =
\4p \int d^2z \sqrt g \ [\    R \l_1
  - \third  R \l_1   + \third  \l_1 \na^2 (2\l_0 - \l_1) - 2 \l_0 \na^2
\l_1   \ ]    \ \ . \eq{A.15} $$
 In the special case when $M_0=M_1=M$ (i.e. $\l_0=\l_1=\l$)  which is relevant
for the discussion of the duality transformation in the path integral
 one has $$
Q_{nm} =    - \d_{nm}\na^2
  -  2( \d_{nm}g^{ab}  + \ep_{nm} \ep^{ab}) \del_a \l \del_b\
 \ \  ,
 \ \  \eq{A.16} $$
 $$  \tQ_{ab}  =  - g_{ab}\na^2 + R_{ab}
  +    2( g_{ab}g^{cd}  + \ep_{ab} \ep^{cd}) \del_c \l \na_d \
 \ \  ,
\eq{A.17} $$ $$ \log Z_Q\vert_{\l_i=\l } \  = \4p \int d^2z \sqrt g \ [\ - {1
\over 12 }  R \na^{-2} R +  R \l  + 2 \del_a \l \del^a \l \ ]   \ \ . \eq{A.18}
$$
It is possible to check the presence of the $\del_a \l \del^a \l $ term in
$\log Z_Q $
by  an  explicit perturbative calculation. Changing the variables from $\k^n$
in
(A.4) to $$u^n = \e{-\l } \k^n$$  for which the scalar product will be
$\l$-independent  one can represent $Z_Q$ as a path integral with the
following action (cf.(A.16))
$$ I= \ha \int d^2z
\sqrt g \
  [  \del^a u^n\del_a u_n \   + 2 \ep_{nm} \ep^{ab}\del_b \l u^n \del_a u^m
  + (\na^2\l + \del_a \l \del^a \l ) u^n u_n  ] \ \ \ ,   $$
or
$$ I = \ha \int d^2z
\sqrt g \
  (  D^a u^n D u_n \
   - F u^n u_n  ) \ \ , $$
$$  D_a u^n = \del_a u^n + A^n_{am}u^m\ \ , \ \ \
A^n_{am}= \ep^n_{\ m }\ep_{a }^{\ b}\del_b \l  \ \ , \ \ \  F\equiv \fourth
\ep^{ab} \ep_{nm} F^{nm}_{ab}= - \na^2 \l \ \ .
$$
Computing the $O(A^2)$ term in the effective action  on a flat background
using dimensional regularisation\foot { It is easy to  check  that UV and IR
divergences cancel separately  so that one  may  set the massless tadpole equal
to zero and use the standard integrals $$ \int {d^dp  \ p_a p_b\over (2\pi)^d
  p^2 (p-q)^2 } =  - {1\over 4 (d-1) } (  q^2 \d_{ab} - d
 q_a q_b ) J\ \ , \ \ \ J= \int {d^dp  \over (2\pi)^d
  p^2 (p-q)^2 } = {1 \over \pi (d-2) } + ... \ \ , $$
as well as $ \ep^{ac}\ep_{ad} = \d^c_d$. } one
finds the finite Schwinger-type term
$$ \log Z_Q =  \4p \int d^2z [ -  \Tr (  A^{\perp}_a )^2 + ... ] \ \  $$
($A^{\perp}_a$ is the transverse part of $A_a$). This result is in
agreement with (A.18).

  2.  The operator equivalent to $Q$  with $M_0=M_1=M$ was
considered in [1,3]  and the presence of the term $\8p \int d^2z \sqrt g \ R\
\log M$  in  the logarithm of its determinant was established by computing the
conformal anomaly. We  have  found  that the complete expression for  $\log
\det
Q$ (A.18) contains also the  $additional$ $M$-dependent term
 $$ \8p \int d^2z \sqrt g \
\del_a \log M  \del^a \log M  \ \ . \eq{A.19}  $$
By  formally doing  the
duality transformation in the path integral one could expect [1,3] that the
ratio (5),(6)  should be  equal to  $Z_1$ (A.1) or to $Z_Q$.
Comparing  (37),(43) with (A.14),(A.15),(A.18) we conclude that  in fact this
cannot   be true  since the term  (A.19) in (A.18)  cannot be present in the
logarithm of (5) (the latter must change sign
 under $M \ra M^{-1}$).

 This apparent contradiction is resolved  by discovering that the
  actual relation between
the torsion (5),(23) and $Z_Q$  contains  also a contribution of
another determinant of second order elliptic operator.
To derive the exact form of this relation let us  repeat the standard steps
corresponding to the duality transformation at the path integral level.\foot
{It should be noted that the discussion which follows is based on operations
with ill-defined integrals and thus is not rigorous.}
 Let us start with the  following path integral
$$ \bZ =
 \int [dp_a] [d \ty ]\exp {( - \int d^2z \sqrt g \ [ \ \ha M(z) g^{ab} p_a p_b
 +  i\ep^{ab} p_a \del_b \ty  ] )} \ \ .
 \eq{A.20} $$
 Integrating  over $p_a$
we get the path integral of the dual theory
$ \bZ \sim  [\det \tD_0]^{-1/2}$ (cf.(5),(6)).
If  one  integrates first over
$\ty$  one obtains the $\d$-function implying that $p^a =  \del_a y$,
i.e. the path integral becomes that of the original theory, i.e. $\bZ \sim
[\det \D_0]^{-1/2} $.  There is, however, an additional determinant which
appears from the $\d$-function.  To  give  a precise sense to the above
relations let us first change the variables as in (A.4)
$$   p_a = \del_a y + \ep_{a}^{ \ b } \del_b \by
 \ \ , \eq{A.21} $$
$$  \bZ =
  J \int [dy] [d\by ] [d \ty ]\exp {( - \int d^2z \sqrt g \ [ \ \ha M(z)
 (\del_a y + \ep_{a}^{ \ b } \del_b \by)^2 +  i  \del^a \by \del_a \ty \ ] )  }
\ \ .
 \eq{A.22} $$
Here $J$ is the Jacobian  corresponding to (A.21) (see (A.5)).
In the context of the sigma model duality  transformation we should assume
that the scalar products in the spaces of $y$ and $\by$  are defined with the
function $M_0=M$ while the scalar product in the space of $\ty$   contains
 $M^{-1}$ (cf.(41),(42)).  Integrating first over $\by$ and $y$ and then
over $\ty$ we get (cf.(A.5))\foot { To compute the integral over  $\by$ and $y$
one is to make a (non-local) shift of the fields which  can be  determined
from the  corresponding shift  of $p_a$ in (A.20). }
 $$ \bZ = \ J \  [\det Q]^{-1/2}   \ [\det \tD_0]^{-1/2} \ \   \ . \eq{A.23} $$
Integrating first over $\ty$ we get the $\d$-function  factor which can be
represented as $\d (\by ) $  multiplied by
 $$ Z_H = \int  [d\by ] [d \ty ] \exp {( - i \int d^2z
\sqrt g \
  \del^a \by \del_a \ty \  )  } \equiv [\det H ]^{-1/2}
\ \ . \eq{A.24} $$
Then
 $$ \bZ = \ J \  [\det H ]^{-1/2} \   [\det \D_0]^{-1/2} \ \   \ . \eq{A.25} $$
Comparing (A.23) with (A.25) we  find  that the following relation must be true
$$
 \det Q  =  \det H \  \det \D_0 \ [\det \tD_0]^{-1}
\ \ , \eq{A.26} $$
i.e.
$$  Z = {\ Z_Q \over  Z_H  } \ \ \ . \eq{A.27} $$
$Z$ and $Z_Q$   defined in
(6),(23) and (A.5) (with $M_i=M$) were already computed in (36) and (A.18).
According to (A.27)
$$  \log Z_H  = \4p \int d^2z \sqrt g \ [\ - {1 \over 12 }  R
\na^{-2} R  +2\del_a \l \del^a \l \ ]   \ \ . \eq{A.28}
$$

 3.   It may be  useful to
 give another  (equivalent)  definition of $Q$  which  makes  the analogy with
the discussion in  Sects.2,3 more transparent.  Given the  DeRham elliptic
complex $(E_i,T_i), i=0,1,2$    one can define the operator
 $$ P: E_0\oplus E_2 \ra E_1 \  \ , \ \ \ \   P (\o_0 ,\ \o_2) = T_0 \o_0 +
T^*_1
\o_2 \ \ ,  $$
which maps a scalar and a 2-form into a vector. Then
$$ P^{\dag } :E_1 \ra E_0\oplus E_2 \ \ , \ \ \ \  P^{\dag} \o_1 = ( T^*_0 \o_1
, \ T_1\o_1) \ \ ,  $$
$$  \P^* P:E_0\oplus E_2\ra E_0\oplus E_2 \ \ , \ \ \
  \P^* P(\o_0 ,\ \o_2) =  (T^*_0T_0\o_0 ,\ T_1T^*_1\o_2) \ \ ,  $$ $$\ \ \
 P \P^* = \D_1=T_0 T_0^* + T^*_1 T_1 \ \ .  $$
 Since  $T_1T_0=0,\ T^*_0 T^*_1 =0$  the operator $ P^{\dag } P$ is diagonal in
$E_0\oplus E_2$ and $\det  P^{\dag } P=\det  P P^{\dag } = \det \D_1$. One can
obtain a new  non-diagonal elliptic second order  operator on $E_0\oplus E_2$
by
introducing a ``twist"  in the definition of $P$,
 $$ P (\o_0 ,\ \o_2) = T_0 \o_0 +  A_1T^*_1A_2 \o_2 \ \ , \eq{A.29} $$
where $A_i$  are  self-adjoint  operators in $E_i$ (e.g. multiplication by
 a  function).  In this case
$$ Q\equiv P^{\dag } P \ \ , \ \ \ Q(\o_0 ,\ \o_2) =
(T^*_0T_0\o_0 + T^*_0 A_1 T^*_1  A_2 \o_2  ,\ A_2 T_1  A_1^2  T^*_1 A_2\o_2 +
A_2 T_1
A_1 T_0 \o_0) \ \ . \eq{A.30} $$
Using  the  relation (17) between $T_i$ and the exterior differentials
$$ T_0=d_0\ \ , \ \ \ T_1=d_1\ \ , \ \ \ T^*_0 =M^{-1}_0d^*_0 M_1 \ \ , \ \ \
 T^*_1 =M^{-1}_1d^*_1 M_2 \ \ , \ \ $$
and comparing (A.4),(A.7) with (A.29),(A.30) we conclude that they are
equivalent  if $$ A_1= M_1 \ \ , \ \ \  A_2= M_2^{-1}\ \ , \ \ \ M_0 = M_2  \
\ , $$ and  if $\o_0$ is identified with $\k^1$ and  the 2-form $ \o_2$ -- with
the scalar $\k^2$.
Explicitly,
$$
 Q=\left(\matrix
{ M^{-1}_0 d^*_0 M_1d_0 \ & \ M^{-1}_0 d^*_0 M_1d^*_1  \cr  M^{-1}_0 d_1 M_1
d_0
\ &\  M_0^{-1} d_1 M_1 d^*_1  \cr}\right)\ \ \ ,
 \eq{A.31} $$
and
$$  \tQ= M_1 PP^{\dag} M_1^{-1} =
  M_1d^*_1 M_0^{-1} d_1  + M_1d_0 M^{-1}_0 d^*_0 \ \ . \ \eq{A.32} $$
Eq.(A.32) is equivalent to (A.11).


\vfill\eject

\centerline{\bf References}
\bigskip

\item {[1]} T.H. Buscher, \pl B194(1987)59 ; \pl B201(1988)466 .
\item {[2]} S. Cecotti, S. Ferrara and L. Girardello, \np B308(1988)436 ;

A. Giveon, E. Rabinovici and G. Veneziano, \np B322(1989)167 ;

G. Veneziano, \pl B265(1991)287 .
\item {[3]} A.A. Tseytlin, \mpl A6(1991)1721 .
\item {[4]} M. Rocek and E. Verlinde, \np B373(1992)630 ;

 A. Giveon and M. Rocek, \np B380(1992)128 .
\item {[5]} B.E. Fridling and A. Jevicki, \pl B134(1984)70 ;

E.S. Fradkin and A.A. Tseytlin, \ap 162(1985)48 .
\item {[6]} P. Ginsparg and C. Vafa, \np B289(1987)414 ;

T. Banks, M. Dine, H. Dijkstra and W. Fischler, \pl B212(1988)466 ;

E. Smith and J. Polchinski, \pl B263(1991)59 .
\item {[7]} P. Gilkey, {\it The Index Theorem and the Heat Equation}
(Publish or Perish, Boston, 1974) .
\item {[8]} A.S. Schwarz, \cmp 67(1979)1 .
\item {[9]} A.S. Schwarz, Lett. Math. Phys. 2(1978)201,247 .
\item {[10]} P.K. Townsend, \pl B88(1979)97 ;

W. Siegel, \pl B93(1979)259 ; \pl B138(1980)107 ;

M.J. Duff and P.van Nieuwenhuizen, \pl B94(1980)179 .
\item {[11]}  A.S. Schwarz and Yu.S. Tyupkin, \np B242(1984)436 .
\item {[12]}   R.T. Seeley, Proc. Symp. Pure Math. 10(1967)288 ;

P. Gilkey, Advances in Math. 10(1973)344 ;
  Proc. Symp. Pure Math. 27(1975)265 ;

Compos. Math. 38(1979)201 .

\vfill \eject
\end


 May be
Eq.(A.28) can be confirmed by the independent computation of $\det H$.
Let us set
$$ \by = \e{\l } ( u^1 + u^2 ) \ \ , \ \ \ty = \e{-\l } ( u^1 - u^2 ) \ \ ,
\ \ \  M=\e{2  \l } \ \ . \eq{A.29} $$
Then the scalar products for  $u^n$  will be  $\l$-independent  so that all
dependence on $\l$ will appear explicitly in
the functional in the exponential in (A.24)
$$ \int d^2z
\sqrt g \
  (  \del^a u^n\del_a u_n \   + 2 \ep_{nm} \del^a \l u^n \del_a u^m
  - \del_a \l \del^a \l u^n u_n  ) \ \ \ , \eq{A.30} $$
$$ u_n  \equiv \eta_{nm} u^m \ \  , \ \ \  \eta_{nm} = diag (1,-1)\ \ . $$
The operator $H$ acting on a pair of scalars  $u^n$  can be represented in the
two equivalent forms
$$ H_{nm} = - \eta_{nm}  \na^2  + 2 \ep_{nm} \del^a \l  \del_a
 - \del_a \l \del^a \l \eta_{nm} \ \ , \eq{A.31} $$
$$ H=
 - \ha  E^{T}
\s \na^2 E \ \ , \eq{A.32} $$ $$ \ E \equiv
 \left(\matrix
{ \e{\l } \ & \ \e{\l }  \cr  \e{ -\l }  \ &\   - \e{-\l }
\cr}\right)
\ \ , \ \ \ \ \s  \equiv \left(\matrix
{ 0 \ & \ 1  \cr  1\ &\   0 \cr}\right) \ \ . $$
As it is clear from (A.32)  the variation of $H$ under the variation of $\l$ is
given by
$$  \d H = \d \l   \s H + H \s  \d \l \ \ . \ \
\eq{A.33} $$
Then (cf.(A.9))
$$ \d \ \log Z_H \ = \ \int d^2z \sqrt g \   \Tr [ (\d \r I - \d \l \s )
{\hat b}_2 (z \vert H) ]  \ \ , \eq{A.34} $$
where $I$ is the unit matrix and ${\hat b}_2$ denotes the matrix value of the
Seeley coefficient, i.e. (27) before taking the trace. Computation of  ${\hat
b}_2 (z \vert H)$ using  (A.31),(26),(27)  gives (cf.(A.12))
$$ {\hat b}_2 (z \vert H)  =
{1\over 4 \pi }\sqrt g \  [ \six I  R     +   \s  \na^2 \l \
]  \ \  \eq{A.35} $$
where
$ \s= (\s^n_m) \ \ , \ \ \ \s^n_{m}= \eta^{nk}\ep_{km} \ \ . $
Integrating (A.34) one  finds  that $Z_H$ is  given by (A.28).